\documentclass[aps,onecolumn,showpacs,nofootinbib]{revtex4}
\usepackage{epsfig}
\usepackage{amsmath}
\usepackage{amsfonts}
\usepackage{amssymb}
\usepackage{graphicx}
\usepackage{colordvi}

\begin{document}

\long\def\symbolfootnote[#1]#2{\begingroup%
\def\thefootnote{\fnsymbol{footnote}}\footnote[#1]{#2}\endgroup}

\title{ The Newtonian  Limit of $f(R)$-gravity}
\author{S. Capozziello\footnote{e\,-\,mail address:
capozziello@na.infn.it}$^{\diamond}$,\,A. Stabile\footnote{e -
mail address: stabile@sa.infn.it}$^{\natural}$, \,A.
Troisi\footnote{e\,-\,mail address:
antrois@gmail.com}$^{\diamond}$}

\affiliation{$^{\diamond}$ Dipartimento di Scienze Fisiche and
INFN, Sez. di Napoli, Universit\`a di Napoli "Federico II",
\\
Compl. Univ. di Monte S. Angelo,
\\Edificio
G, Via Cinthia, I-80126 - Napoli, Italy }

\affiliation{$^{\natural}$ Dipartimento di Fisica "E. R.
Caianiello", Universita' degli Studi di Salerno,\\
 Via S. Allende, I-84081 Baronissi (SA), Italy.}
\begin{abstract}
A general analytic procedure is developed  to deal with the
Newtonian  limit of  $f(R)$ gravity. A discussion comparing the
Newtonian and the post-Newtonian limit of these models is proposed
in order to point out the differences between the two approaches.
We calculate the post-Newtonian parameters of such theories
without any redefinition of the degrees of freedom, in particular,
without adopting  some scalar fields and without any change from
Jordan to Einstein frame. Considering the Taylor expansion of a
generic $f(R)$ theory, it is possible to obtain general solutions
in term of the metric coefficients up to the third order of
approximation. In particular, the solution relative to the
$g_{tt}$ component gives a gravitational potential always
corrected with respect to the Newtonian one of the linear theory
$f(R)=R$. Furthermore, we show that the Birkhoff theorem is not a
general result for $f(R)$-gravity since time-dependent evolution
for spherically symmetric solutions can be achieved depending on
the order of perturbations. Finally, we discuss the
post-Minkowskian limit and the emergence of massive gravitational
wave solutions.
\end{abstract}
 \maketitle
\section{Introduction}

In  recent years, the effort to give a  physical explanation to
the today observed cosmic acceleration \cite{sneIa,lss,cmbr} has
attracted a good amount of interest in $f(R)$-gravity, considered
as a viable mechanism to explain the cosmic acceleration by
extending the geometric sector of field equations
\cite{f(R)-noi,f(R)-cosmo,palatini}. There are several physical
and mathematical motivations to enlarge General Relativity (GR) by
these theories. For  comprehensive review, see
\cite{GRGreview,OdintsovLadek,farhoudi}.

Specifically, cosmological models coming from $f(R)$-gravity were
firstly introduced by Starobinsky \cite{starobinsky} in the early
80'ies to build up a feasible inflationary model where geometric
degrees of freedom had the role of the scalar field ruling the
inflation and the structure formation.

On the other side, dealing with such  extended gravity models at
shorter astrophysical scales (Galaxy and Solar System), one faces
the emergence of corrected gravitational potentials with respect
to the Newton one coming out from GR. This result is well known
since a long time \cite{stelle}, and recently it has been pursued
to carry out the possibility to explain the flatness of spiral
galaxies rotation curves  without the addition of huge amount of
Dark Matter. In particular, the rotation curves of a wide sample
of low-surface-brightness spiral galaxies have been successfully
fitted by these corrected potentials \cite{noi-mnras} and reliable
results are also expected for other galaxy-types \cite{salucci}.

Other issues as, for example, the observed Pioneer anomaly problem
\cite{anderson} can be framed into the same approach
\cite{bertolami} and then, apart the cosmological dynamics, a
systematic analysis of such theories urges at short scale and in
the low energy limit.

In this paper, we are going to discuss,  without specifying the
form of the theory, the Newtonian limit of $f(R)$-gravity pointing
out the differences and the relations with respect the
post-Newtonian  and the post-Minkowskian limits. In literature,
there are several definitions and  several claims in this
direction but clear statements and discussion on these approaches
urge in order to find out definite results to be tested by
experiments \cite{faraonithomas}.

The discussion about the short scale behavior of higher order
gravity has been quite vivacious in the last years since  GR shows
is best predictions just at the Solar System level. As matter of
fact, measurements coming from weak field limit tests like the
bending of light, the perihelion shift of planets, frame dragging
experiments represent inescapable tests for whatever theory of
gravity. Actually, in our opinion, there are sufficient
theoretical predictions to state that higher order theories of
gravity can be compatible with Newtonian and post-Newtonian
prescriptions. In other papers \cite{ppn-noi}, we have shown that
this result can be achieved by means of the analogy of
$f(R)$\,-\,models with scalar\,-\,tensor gravity.

Nevertheless, up to now, the discussion on the weak field limit of
$f(R)$\,-\,theories is far to be definitive and there are several
papers claiming for opposite results \cite{ppn-ok,ppn-no}, or
stating that no progress has been reached in the last  forty due
to the several common misconceptions in the various theories of
gravity \cite{faraonithomas}.

In particular, people approached the weak limit issue following
different schemes and developing different parameterizations
which, in some cases, turn out to be not necessarily correct.

The purpose is to take part to the debate, building up a rigorous
formalism which deals with the formal definition of weak field and
small velocities limit applied to fourth-order gravity. In a
series of papers, our aim is to pursue a systematic discussion
involving: $i)$ the Newtonian limit of  $f(R)$-gravity (the
present paper), $ii)$ spherically symmetric solutions vs. the weak
filed limit in $f(R)$ -gravity \cite{artspher}; and, finally,
$iii)$ general fourth-order theories where also invariants as
$R_{\mu\nu}R^{\mu\nu}$ or
$R_{\alpha\beta\mu\nu}R^{\alpha\beta\mu\nu}$ are considered,
\cite{dirk}.

Our analysis is based on the metric approach, developed in the
Jordan frame, assuming that the observations are performed in it,
without resorting to any conformal transformation as done in
several cases \cite{olmo}. This point of view is adopted in order
to avoid dangerous variable changes which could compromise the
correct physical interpretation of the results.

We will show that the corrections induced on the gravitational
potentials can be suitable to explain relevant astrophysical
behaviors or can be related with some relevant physical issues.

As a preliminary analysis,  we will  concentrate on the vacuum
case with the aim to build up a further rigorous formalism for the
Newtonian and post-Newtonian limit of $f(R)$ theories in presence
of matter. As we will see, it is possible to deduce an effective
estimation   of the post-Newtonian parameter $\gamma$ by
considering the second order solutions of the metric coefficient
in the vacuum case. For the sake of completeness we will treat the
problem also by imposing the harmonic gauge on the field
equations.
\\
The paper is organized as follows: in Sec. II,  the general
formalism concerning the spherically symmetric background in
fourth order gravity is introduced; Sec. III is devoted to a
discussion of the post-Newtonian approximation considering the
differences with respect GR: in this theory not all order of
perturbations can be consistently achieved if conservation laws
are taken into account, in $f(R)$-gravity this shortcoming can be,
in principle, avoided. In Sec. IV, the analytic approach to the
weak field in $f(R)$-gravity is developed. In particular, we
achieve the  gravitational potential (related to the
$g_{tt}$-component of the metric)  which is always corrected with
respect to the Newtonian one of the linear $f(R)=R$ theory.
Besides, we show that the Birkhoff theorem is not a general result
for $f(R)$-gravity since time-dependent evolution for spherically
symmetric solutions can be achieved depending on the order of
perturbations. In Sec. V, the post-Minkowskian limit is discussed
considering also the possibility to obtain gravitational waves
solutions. Sec. VI is devoted to the discussion and conclusions.

\section{$f(R)$\,-\,gravity in spherically symmetric spacetime}

The  action for $f(R)$- gravity  reads\,:

\begin{equation}\label{actfR}
{\cal A}\,=\, \int
d^4x\sqrt{-g}\biggl[f(R)+\mathcal{X}\mathcal{L}_m\biggr]\,,
\end{equation}
where $f(R)$ is an  analytic function of Ricci scalar,
$\mathcal{X}=\frac{16\pi G}{c^4}$ is the coupling constant and
$\mathcal{L}_m$ describes the ordinary matter Lagrangian. Such an
action is the straightforward generalization of the
Hilbert-Einstein action of GR where $f(R)=R$ is assumed.

By varying  (\ref{actfR}) with respect to the metric, one obtains
the fourth-order field equations\,:

\begin{equation}\label{fe}
f'R_{\mu\nu}-\frac{1}{2}fg_{\mu\nu}-f'_{;\mu\nu}+g_{\mu\nu}\Box
f'=\frac{\mathcal{X}}{2}T_{\mu\nu}\,,
\end{equation}
with ${\displaystyle
T_{\mu\nu}=\frac{-2}{\sqrt{-g}}\frac{\delta(\sqrt{-g}\mathcal{L}_m)}{\delta
g^{\mu\nu}}}$ and ${\displaystyle f'=\frac{df(R)}{dR}}$. The trace
 is
\begin{equation}
3\Box f'+f'R-2f=\frac{\mathcal{X}}{2}T\,,
\end{equation}
and such an expression can be read as a Klein-Gordon equation,
where the effective field is $f'$, if $f(R)$ is non-linear in $R$
\cite{starobinsky}.

As said,  we are interested in investigating the Newtonian and the
post-Newtonian limit of $f(R)$-gravity in a spherically symmetric
background.  Solutions   can be obtained considering the metric
(see also \cite{noether,multamaki})\,:

\begin{equation}\label{me}
ds^2\,=\,g_{\sigma\tau}dx^\sigma
dx^\tau=A(x^0,r)d{x^0}^2-B(x^0,r)dr^2-r^2d\Omega
\end{equation}
where $x^0\,=\,ct$; $A$ and $B$ are generic functions depending on
time and  coordinate radius; $d\Omega$ is the  angular element.
The field equations (\ref{fe}) turn out to be

\begin{equation}\label{fe1}
\begin{array}{ll}
H_{\mu\nu}=f'R_{\mu\nu}-\frac{1}{2}fg_{\mu\nu}+\mathcal{H}_{\mu\nu}=\frac{\mathcal{X}}{2}T_{\mu\nu}
\\\\
H=g^{\sigma\tau}H_{\sigma\tau}=f'R-2f+\mathcal{H}=\frac{\mathcal{X}}{2}T
\end{array}
\end{equation}
where

\begin{equation}
\begin{array}{ll}
\mathcal{H}_{\mu\nu}=-f''\biggl\{R_{,\mu\nu}-\Gamma^0_{\mu\nu}R_{,0}-\Gamma^r_{\mu\nu}R_{,r}-
g_{\mu\nu}\biggl[\biggl({g^{00}}_{,0}+g^{00}
\ln\sqrt{-g}_{,0}\biggr)R_{,0}+\biggl({g^{rr}}_{,r}+g^{rr}\ln\sqrt{-g}_{,r}\biggr)R_{,r}+\\\\+g^{00}R_{,00}
+g^{rr}R_{,rr}\biggr]\biggr\}-f'''\biggl[R_{,\mu}R_{,\nu}-g_{\mu\nu}\biggl(g^{00}{R_{,0}}^2+g^{rr}
{R_{,r}}^2\biggr)\biggr]
\\\\
\mathcal{H}=g^{\sigma\tau}\mathcal{H}_{\sigma\tau}=3f''\biggl[\biggl({g^{00}}_{,0}+g^{00}
\ln\sqrt{-g}_{,0}\biggr)R_{,0}+\biggl({g^{rr}}_{,r}+g^{rr}\ln\sqrt{-g}_{,r}\biggr)R_{,r}+g^{00}R_{,00}
+g^{rr}R_{,rr}\biggr]+\\\\+
3f'''\biggl[g^{00}{R_{,0}}^2+g^{rr}{R_{,r}}^2\biggr]
\end{array}
\end{equation}
are the higher than second order terms of the theory. We are
adopting  the convention $R_{\mu\nu}={R^\rho}_{\mu\rho\nu}$ for
the Ricci tensor and
${R^\alpha}_{\beta\mu\nu}=\Gamma^\alpha_{\beta\nu,\mu}-...$, for
the Riemann tensor. Connections are Levi-Civita \,:

\begin{equation}\label{chri}
\Gamma^\mu_{\alpha\beta}=\frac{1}{2}g^{\mu\rho}(g_{\alpha\rho,\beta}+g_{\beta\rho,\alpha}-g_{\alpha\beta,\rho})\,.
\end{equation}

\section{General remarks on the Newtonian and the post-Newtonian approximation}

At this point, it is worth discussing  some general issues on the
Newtonian and post-Newtonian limits. Basically there are some
general features one has to take into account when approaching
these limits, whatever the underlying theory of gravitation is.

If one consider a system of gravitationally interacting particles
of  mass $\bar{M}$, the kinetic energy
$\frac{1}{2}\bar{M}\bar{v}^2$ will be, roughly, of the same order
of magnitude as the typical potential energy
$U=G\bar{M}^2/\bar{r}$, with $\bar{M}$, $\bar{r}$, and $\bar{v}$
the typical average values of  masses, separations, and velocities
of these particles. As a consequence:

\begin{equation}
\bar{v^2}\sim \frac{G\bar{M}}{\bar{r}}\,,
\end{equation}
(for instance, a test particle in a circular orbit of radius $r$
about a central mass $M$ will have velocity $v$ given in Newtonian
mechanics by the exact formula $v^2=GM/r$.)

The post-Newtonian approximation can be described as a method for
obtaining the motion of the system to an higher than the first
order (approximation which coincides with the Newtonian mechanics)
with respect to the quantities $G\bar{M}/\bar{r}$ and $\bar{v}^2$
assumed small with respect to the squared light speed $c^2$. This
approximation is sometimes referred to as an expansion in inverse
powers of the light speed.

The typical values of the Newtonian gravitational potential $U$
are nowhere larger than $10^{-5}$ in the Solar System (in
geometrized units, $U/c^2$ is dimensionless). On the other hand,
planetary velocities satisfy the condition $\bar{v}^2\lesssim U$,
while\symbolfootnote[4]{We consider here on the velocity $v$ in
units of the light speed $c$.} the matter pressure $p$ experienced
inside the Sun and the planets is generally smaller than the
matter gravitational energy density $\rho U$, in other words
\symbolfootnote[3]{Typical values of $p/\rho$ are $\sim 10^{-5}$
in the Sun and  $\sim 10^{-10}$ in the Earth \cite{will}.}
$p/\rho\lesssim U$. Furthermore one must consider that even other
forms of energy in the Solar System (compressional energy,
radiation, thermal energy, etc.) have small intensities and the
specific energy density $\Pi$ (the ratio of the energy density to
the rest-mass density) is related to $U$ by $\Pi\lesssim U$ ($\Pi$
is $\sim 10^{-5}$ in the Sun and $\sim 10^{-9}$ in the Earth
\cite{will}). As matter of fact, one can consider that these
quantities, as function of the velocity, give second order
contributions\,:

\begin{equation}
U\sim v^2\sim p/\rho\sim \Pi\sim\text{O(2)}\,.
\end{equation}
Therefore, the velocity $v$ gives O(1) terms in the velocity
expansions, $U^2$ is of order O(4), $Uv$ of O(3), $U\Pi$ is of
O(4), and so on. Considering these approximations, one has

\begin{equation}
\frac{\partial}{\partial x^0}\sim\textbf{v}\cdot\nabla\,,
\end{equation}
and

\begin{equation}
\frac{|\partial/\partial x^0|}{|\nabla|}\sim\text{O(1)}\,.
\end{equation}
Now,  particles move along geodesics\,:

\begin{equation}
\frac{d^2x^\mu}{ds^2}+\Gamma^\mu_{\sigma\tau}\frac{dx^\sigma}{ds}\frac{dx^\tau}{ds}=0\,,
\end{equation}
which can be written in details as

\begin{equation}
\frac{d^2x^i}{dx^{0\,2}}=-\Gamma^i_{00}-2\Gamma^i_{0m}\frac{dx^m}{dx^0}-
\Gamma^i_{mn}\frac{dx^m}{dx^0}\frac{dx^n}{dx^0}+\biggl[\Gamma^0_{00}+
2\Gamma^0_{0m}\frac{dx^m}{dx^0}+2\Gamma^0_{mn}\frac{dx^m}{dx^0}\frac{dx^n}{dx^0}\biggr]\frac{dx^i}{dx^0}\,.
\end{equation}
In the Newtonian approximation, that is vanishingly small
velocities and only  first-order terms in the difference between
$g_{\mu\nu}$ and the Minkowski metric $\eta_{\mu\nu}$, one obtains
that the particle motion equations reduce to the standard
result\,:

\begin{equation}
\frac{d^2x^i}{dx^{0\
2}}\simeq-\Gamma^i_{00}\simeq-\frac{1}{2}\frac{\partial
g_{00}}{\partial x^i}\,.
\end{equation}
The quantity $1-g_{00}$ is of order $G\bar{M}/\bar{r}$, so that
the Newtonian approximation gives
$\displaystyle\frac{d^2x^i}{dx^{0\,2}}$ to the order
$G\bar{M}/\bar{r}^2$, that is, to the order $\bar{v}^2/r$. As a
consequence if we would like to search for the post-Newtonian
approximation, we need to compute
$\displaystyle\frac{d^2x^i}{dx^{0\,2}}$ to the order
$\bar{v}^4/\bar{r}$. Due to the Equivalence Principle and the
differentiability of spacetime manifold, we expect that it should
be possible to find out a coordinate system in which the metric
tensor is nearly equal to the Minkowski one $\eta_{\mu\nu}$, the
correction being expandable in powers of
$G\bar{M}/\bar{r}\sim\bar{v}^2$. In other words one has to
consider the metric developed as follows\,:

\begin{equation}\label{approx1}
\left\{\begin{array}{ll}g_{00}(x^0,\textbf{x})\simeq1+g^{(2)}_{00}(x^0,\textbf{x})+g^{(4)}_{00}(x^0,\textbf{x})+\text{O(6)}
\\\\g_{0i}(x^0,\textbf{x})\simeq g^{(3)}_{0i}(x^0,\textbf{x})+\text{O(5)}\\\\
g_{ij}(x^0, \textbf{x})\simeq-\delta_{ij}+g^{(2)}_{ij}(x^0,
\textbf{x})+\text{O(4)}
\end{array}\right.
\end{equation}
where $\delta_{ij}$ is the Kronecker delta, and for the
controvariant form of $g_{\mu\nu}$, one has

\begin{equation}\label{approx2}
\left\{\begin{array}{ll}g^{00}(x^0,\textbf{x})\simeq
1+g^{(2)00}(x^0, \textbf{x})+g^{(4)00}(x^0,
\textbf{x})+\text{O(6)}
\\\\
g^{0i}(x^0,\textbf{x})\simeq g^{(3)0i}(x^0,\textbf{x})+\text{O(5)}\\\\
g^{ij}(x^0,\textbf{x})\simeq-\delta^{ij}+g^{(2)ij}(x^0,{\textbf{x}})+{\text{O(4)}}\,.
\end{array}\right.
\end{equation}
In evaluating $\Gamma^\mu_{\alpha\beta}$ we must take into account
that the scale of distance and time, in our systems, are
respectively set by $\bar{r}$ and $\bar{r}/\bar{v}$, thus the
space and time derivatives should be regarded as being of order

\begin{equation}
\frac{\partial}{\partial x^i}\sim\frac{1}{\bar{r}}\,, \ \ \ \ \ \
\ \frac{\partial}{\partial x^0}\sim\frac{\bar{v}}{\bar{r}}\,.
\end{equation}
Using the above approximations (\ref{approx1}), (\ref{approx2}),
we have, from the definition (\ref{chri}),

\begin{equation}
\left\{\begin{array}{ll} \begin{array}{ccc}
  {\Gamma^{(3)}}^{0}_{00}=\frac{1}{2}g^{(2),0}_{00}\, & \,\,\, & {\Gamma^{(2)}}^{i}_{00}=\frac{1}{2}g^{(2),i}_{00} \\
  & & \\
{\Gamma^{(2)}}^{i}_{jk}=\frac{1}{2}\biggl({g^{(2)}}^{,i}_{jk}-{g^{(2)}}^{i}_{j,k}-{g^{(2)}}^{i}_{k,j}\biggr)\,
& \,\,\, & {\Gamma^{(3)}}^{0}_{ij}=\frac{1}{2}\biggl
  ({g^{(3)}}^{0}_{i,j}+{g^{(3)}}^{0}_{j,i}-{g^{(3)}}^{,0}_{ij}\biggr) \\
  & & \\
{\Gamma^{(3)}}^{i}_{0j}=\frac{1}{2}\biggl({g^{(3)}}^{,i}_{0j}-{g^{(3)}}^{i}_{0,j}-{g^{(2)}}^{i}_{j,0}\biggr)\,
& \,\,\, & {\Gamma^{(4)}}^{0}_{0i}=\frac{1}{2}\biggl
  ({g^{(4)}}^{0}_{0,i}+g^{(2)00}g^{(2)}_{00,i}\biggr) \\
  & & \\
  {\Gamma^{(4)}}^{i}_{00}=\frac{1}{2}\biggl({g^{(4)}}^{,i}_{00}+g^{(2)im}g^{(2)}_{00,m}-2{g^{(3)}}^{i}_{0,0}\biggr)\, & \,\,\, &
  {\Gamma^{(2)}}^{0}_{0i}= \frac{1}{2}{g^{(2)}}^{0}_{0,i}
  \end{array}
\end{array}\right.
\end{equation}
The Ricci tensor component are

\begin{equation}
\left\{\begin{array}{ll}R^{(2)}_{00}=\frac{1}{2}\triangle
g^{(2)}_{00}\\\\R^{(4)}_{00}=\frac{1}{2}\triangle
g^{(4)}_{00}-\frac{1}{2}{g^{(2)}}^{mn}_{,m}g^{(2)}_{00,n}-\frac{1}{2}{g^{(2)}}^{mn}g^{(2)}_{00,mn}+\frac{1}{2}{g^{(2)}}^{m}_{m,00}-\frac{1}{4}{g^{(2)}}^{0,m}_{0}g^{(2)}_
{00,m}-\frac{1}{4}{g^{(2)}}^{m,n}_{m}g^{(2)}_{00,n}-{g^{(3)}}^{m}_{0,m0}\\\\R^{(3)}_{0i}=\frac{1}{2}\triangle
g^{(3)}_{0i}-\frac{1}{2}{g^{(2)}}^{m}_{i,m0}-\frac{1}{2}{g^{(3)}}^{m}_{0,mi}+\frac{1}{2}{g^{(2)}}^{m}_{m,0i}\\\\R^{(2)}_{ij}=\frac{1}{2}\triangle
g^{(2)}_{ij}-\frac{1}{2}{g^{(2)}}^{m}_{i,mj}-\frac{1}{2}{g^{(2)}}^{m}_{j,mi}-\frac{1}{2}{g^{(2)}}^{0}_{0,ij}+\frac{1}{2}{g^{(2)}}^{m}_{m,ij}\end{array}\right.
\end{equation}
and assuming the harmonic gauge
$g^{\rho\sigma}\Gamma^\mu_{\rho\sigma}=0$ (see the Appendix for
details), one can rewrite these last expressions as

\begin{equation}\label{PPNobj}
\left\{\begin{array}{ll}R^{(2)}_{00}=\frac{1}{2}\triangle
g^{(2)}_{00}\\\\R^{(4)}_{00}=\frac{1}{2}\triangle
g^{(4)}_{00}-\frac{1}{2}{g^{(2)}}^{mn}g^{(2)}_{00,mn}-\frac{1}{2}{g^{(2)}}^{0}_{0,00}-\frac{1}{2}|\bigtriangledown_\eta
g^{(2)}_{00}|^2\\\\R^{(3)}_{0i}=\frac{1}{2} \triangle
g^{(3)}_{0i}\\\\R^{(2)}_{ij}=\frac{1}{2}\triangle
g^{(2)}_{ij}\end{array}\right.
\end{equation}
with $\triangle$ and $\bigtriangledown$, respectively, the
Laplacian and the gradient in flat space. The Ricci scalar reads

\begin{equation}
\left\{\begin{array}{ll}R^{(2)}={R^{(2)}}^{0}_{0}-{R^{(2)}}^{m}_{m}=\frac{1}{2}\triangle{g^{(2)}}^{0}_{0}-\frac{1}{2}
\triangle{g^{(2)}}^{m}_{m}\\\\R^{(4)}={R^{(4)}}^{0}_{0}+{g^{(2)}}^{00}
R^{(2)}_{00}+{g^{(2)}}^{mn}R^{(2)}_{mn}=\frac{1}{2}\triangle
{g^{(4)}}^{0}_{0}-\frac{1}{2}{g^{(2)}}^{0,0}_{0,0}-\frac{1}{2}{g^{(2)}}^{mn}\biggl[{g^{(2)}}^{0}_{0,mn}-
\triangle g^{(2)}_{mn}\biggr]-\frac{1}{2}|\bigtriangledown
{g^{(2)}}^{0}_{0}|^2+\frac{1}{2}{g^{(2)}}^{00}\triangle
g^{(2)}_{00}\end{array}\right.\,.
\end{equation}
The inverse of the metric tensor is defined by means of the
equation

\begin{equation}
g^{\alpha\rho}g_{\rho\beta}=\delta^\alpha_\beta
\end{equation}
with $\delta^\alpha_\beta$ the Kronecker delta. The relations
among the higher than first order terms turn out to be

\begin{equation}
\left\{\begin{array}{ll}g^{(2)00}(x_0,\textbf{x})=-g^{(2)}_{00}(x_0,\textbf{x})\\\\g^{(4)00}(x_0,\textbf{x})={g^{(2)}_{00}(x_0,\textbf{x})}^2-g^{(4)}_{00}
(x_0,\textbf{x})\\\\g^{(3)0i}=g^{(3)}_{0i}\\\\g^{(2)ij}(x_0,\textbf{x})=-g^{(2)}_{ij}(x_0,\textbf{x})
\end{array}\right.
\end{equation}
Finally the Lagrangian of a particle in presence of a
gravitational field can be expressed as proportional to the
invariant distance $ds^{1/2}$, thus we have\,:

\begin{equation}
L=\biggl(g_{\rho\sigma}\frac{dx^\rho}{dx^0}\frac{dx^\sigma}{dx^0}\biggr)^{1/2}=\biggl(g_{00}+2g_{0m}v^m+g_{mn}v^mv^n\biggr)^{1/2}=
\biggl(1+g^{(2)}_{00}+g^{(4)}_{00}+2g^{(3)}_{0m}v^m-\textbf{v}^2+g^{(2)}_{mn}v^mv^n\biggr)^{1/2}\,,
\end{equation}
which, to the O(2) order, reduces to the classic Newtonian
Lagrangian of a test particle
$L_{\text{New}}=\biggl(1+g^{(2)}_{00}-\textbf{v}^2\biggr)^{1/2}$,
where $\mathbf{v}=\frac{dx^m}{dx^0}\frac{dx_m}{dx^0}$. As matter
of fact, post-Newtonian physics has to involve higher than O(4)
order terms in the Lagrangian.

An important remark concerns the odd-order perturbation terms O(1)
or O(3). Since, these terms contain  odd powers of velocity
$\textbf{v}$ or of time derivatives, they are related to the
energy dissipation or absorption by the system. Nevertheless, the
mass-energy conservation prevents the energy and mass losses and,
as a consequence, prevents, in the Newtonian limit, terms of O(1)
and O(3) orders in the  Lagrangian. If one takes into account
contributions  higher than O(4) order, different theories give
different predictions. GR, for example, due to the conservation of
post-Newtonian energy, forbids  terms of O(5) order; on the other
hand, terms of O(7) order can appear and are related to the energy
lost by means of the gravitational radiation.

\section{The Newtonian limit of $f(R)$ gravity in  spherically symmetric background vs. post-Newtonian limit}

Exploiting the formalism of post-Newtonian approximation described
in the previous section, we can  develop  a systematic analysis in
the limit of weak field and small velocities for $f(R)$-gravity.
We are going to assume, as background,  a spherically symmetric
spacetime  and we are going to investigate the vacuum case.
Considering the metric (\ref{me}), assuming, unless not specified,
$c\,=\,1$ and then $x^0=ct\rightarrow t$, we have, for a given
$g_{\mu\nu}$\,:

\begin{equation}\label{definexpans}
\left\{\begin{array}{ll} g_{tt}(t,
r)=A(t,r)\simeq1+g^{(2)}_{tt}(t,r)+g^{(4)}_{tt}(t,r)
\\\\
g_{rr}(t,r)=-B(t,r)\simeq-1+g^{(2)}_{rr}(t,r)\\\\
g_{\theta\theta}(t,r)=-r^2\\\\
g_{\phi\phi}(t,r)=-r^2\sin^2\theta
\end{array}\right.\,,
\end{equation}
while the approximations for $g^{\mu\nu}$ are

\begin{equation}
\left\{\begin{array}{ll} g^{tt}=A(t,r)^{-1}\simeq
1-g^{(2)}_{tt}+[{g^{(2)}_{tt}}^2-g^{(4)}_{tt}]
\\\\
g^{rr}=-B(t,r)^{-1}\simeq-1-g^{(2)}_{rr}
\end{array} \right.\,.
\end{equation}
The determinant reads

\begin{equation}
g\simeq
r^4\sin^2\theta\{-1+[g^{(2)}_{rr}-g^{(2)}_{tt}]+[g^{(2)}_{tt}g^{(2)}_{rr}-g^{(4)}_{tt}]\}\,.
\end{equation}
As a consequence, the Christoffel's symbols are

\begin{equation}
\left\{\begin{array}{ll} \begin{array}{ccc}
  {\Gamma^{(3)}}^{t}_{tt}=\frac{g^{(2)}_{tt,t}}{2}\, & \,\,\, & {\Gamma^{(2)}}^{r}_{tt}+{\Gamma^{(4)}}^{r}_{tt}=
  \frac{g^{(2)}_{tt,r}}{2}+\frac{g^{(2)}_{rr}g^{(2)}_{tt,r}+g^{(4)}_{tt,r}}{2} \\
  & & \\
  {\Gamma^{(3)}}^{r}_{tr}=-\frac{g^{(2)}_{rr,t}}{2}\, & \,\,\, & {\Gamma^{(2)}}^{t}_{tr}+{\Gamma^{(4)}}^{t}_{tr}
  =\frac{g^{(2)}_{tt,r}}{2}+\frac{g^{(4)}_{tt,r}-g^{(2)}_{tt}g^{(2)}_{tt,r}}{2} \\
  & & \\
  {\Gamma^{(3)}}^{t}_{rr}=-\frac{g^{(2)}_{rr,t}}{2}\, & \,\,\, & {\Gamma^{(2)}}^{r}_{rr}+{\Gamma^{(4)}}^{r}_{rr}
  =-\frac{g^{(2)}_{rr,r}}{2}-\frac{g^{(2)}_{rr}g^{(2)}_{rr,r}}{2} \\
  & & \\
  \Gamma^{r}_{\phi\phi}=\sin^2\theta \Gamma^{r}_{\theta\theta}\, & \,\,\, & {\Gamma^{(0)}}^{r}_{\theta\theta}+
  {\Gamma^{(2)}}^{r}_{\theta\theta}+{\Gamma^{(4)}}^{r}_{\theta\theta}=-r-rg^{(2)}_{rr}-r{g^{(2)}_{rr}}^2 \\
\end{array}
\end{array} \right.
\end{equation}
Let us even display the Ricci's tensor components

\begin{equation}
\left\{\begin{array}{ll} R_{tt}&\simeq R^{(2)}_{tt}+R^{(4)}_{tt}
\\\\
R_{tr}&\simeq R^{(3)}_{tr}\\\\
R_{rr}&\simeq R^{(2)}_{rr}\\\\
R_{\theta\theta}&\simeq R^{(2)}_{\theta\theta}\\\\
R_{\phi\phi}&\simeq\sin^2\theta R^{(2)}_{\theta\theta}
\end{array} \right.
\end{equation}
where

\begin{equation}
\left\{\begin{array}{ll}
R^{(2)}_{tt}=\frac{rg^{(2)}_{tt,rr}+2g^{(2)}_{tt,r}}{2r}
\\\\R^{(4)}_{tt}=\frac{-r{g^{(2)}_{tt,r}}^2+4g^{(4)}_{tt,r}+rg^{(2)}_{tt,r}g^{(2)}_{rr,r}+2g^{(2)}_{rr}[2g^{(2)}_{tt,r}+rg^{(2)}_{tt,rr}]+2rg^{(4)}_{tt,rr}
+2rg^{(2)}_{rr,tt}}{4r}\\\\
R^{(3)}_{tr}=-\frac{g^{(2)}_{rr,t}}{r}\\\\
R^{(2)}_{rr}=-\frac{rg^{(2)}_{tt,rr}+2g^{(2)}_{rr,r}}{2r}\\\\
R^{(2)}_{\theta\theta}=-\frac{2g^{(2)}_{rr}+r[g^{(2)}_{tt,r}+g^{(2)}_{rr,r}]}{2}
\end{array} \right.
\end{equation}
and the Ricci  scalar expression in the post-Newtonian
approximation

\begin{equation}R\simeq R^{(2)}+R^{(4)}\end{equation}
with

\begin{equation}
\left\{\begin{array}{ll}
R^{(2)}=\frac{2g^{(2)}_{rr}+r[2g^{(2)}_{tt,r}+2g^{(2)}_{rr,r}+rg^{(2)}_{tt,rr}]}{r^2}
\\\\
R^{(4)}=\frac{4{g^{(2)}_{rr}}^2+2rg^{(2)}_{rr}[2g^{(2)}_{tt,r}+4g^{(2)}_{rr,r}+rg^{(2)}_{tt,rr}]+r\{-r{g^{(2)}_{tt,r}}^2+4g^{(4)}_{tt,r}+r
g^{(2)}_{tt,r}g^{(2)}_{rr,r}-2g^{(2)}_{tt}[2g^{(2)}_{tt,r}+rg^{(2)}_{tt,rr}]+2rg^{(4)}_{tt,rr}+2rg^{(2)}_{rr,tt}\}}{2r^2}
\end{array} \right.\,.
\end{equation}
In order to derive the post-Newtonian approximation for a generic
function $f(R)$,  one should specify the $f(R)$\,-\,Lagrangian
into the field equations (\ref{fe1}). This is a crucial point
because once a certain Lagrangian is chosen, one will obtain a
particular post-Newtonian approximation referred to such a choice.
This means to lose any general prescription and to obtain
corrections to the Newtonian potential which refer "univocally" to
the considered $f(R)$ function. Alternatively, one can restrict to
analytic $f(R)$ functions   expandable with respect to a certain
value $R\,=\,R_0$. In general, such theories are physically
interesting and allow to recover the GR  results and the correct
boundary and asymptotic conditions. Then we assume

\begin{equation}\label{sertay}
f(R)=\sum_{n}\frac{f^n(R_0)}{n!}(R-R_0)^n\simeq
f_0+f_1R+f_2R^2+f_3R^3+...\,.
\end{equation}
On the other hand,  it is possible to obtain the post-Newtonian
approximation of $f(R)$-gravity considering such an expansion into
the field equations (\ref{fe1}) and expanding the system up to the
orders O(0), O(2) e O(4). This approach  provides  general results
and specific (analytic) Lagrangians are selected by  the
coefficients $f_i$ in (\ref{sertay}).

Let us now substitute the series (\ref{sertay}) into the field
Eqs. (\ref{fe1}). Developing the equations up to O(0), O(2) and
O(4) orders in the case of vanishing matter, i.e. $T_{\mu\nu}=0$,
we have

\begin{equation}\label{sys1}
\left\{\begin{array}{ll} H^{(0)}_{\mu\nu}=0,\,\,&\,\,H^{(0)}=0
\\\\
H^{(2)}_{\mu\nu}=0,\,\,&\,\,H^{(2)}=0\\\\
H^{(3)}_{\mu\nu}=0,\,\,&\,\,H^{(3)}=0\\\\
H^{(4)}_{\mu\nu}=0,\,\,&\,\,H^{(4)}=0\\\\
\end{array} \right.
\end{equation}
and, in particular, from the O(0) order approximation, one obtains

\begin{equation}\label{eq0}
f_0=0\,.
\end{equation}
This result suggests a first  consideration. If the Lagrangian is
developable around a vanishing value of the Ricci scalar ($R_0=0$)
the relation (\ref{eq0}) will imply that the cosmological constant
contribution has to be zero in vacuum whatever is the
$f(R)$-gravity theory. This result appears quite obvious but
sometime it is not considered in literature.

If we now consider the O(2)- order approximation, the equations
system (\ref{sys1}), in the vacuum case, results to be
\begin{equation}\label{eq2}
\left\{\begin{array}{ll}
f_1rR^{(2)}-2f_1g^{(2)}_{tt,r}+8f_2R^{(2)}_{,r}-f_1rg^{(2)}_{tt,rr}+4f_2rR^{(2)}=0
\\\\
f_1rR^{(2)}-2f_1g^{(2)}_{rr,r}+8f_2R^{(2)}_{,r}-f_1rg^{(2)}_{tt,rr}=0
\\\\
2f_1g^{(2)}_{rr}-r[f_1rR^{(2)}-f_1g^{(2)}_{tt,r}-f_1g^{(2)}_{rr,r}+4f_2R^{(2)}_{,r}+4f_2rR^{(2)}_{,rr}]=0
\\\\
f_1rR^{(2)}+6f_2[2R^{(2)}_{,r}+rR^{(2)}_{,rr}]=0
\\\\
2g^{(2)}_{rr}+r[2g^{(2)}_{tt,r}-rR^{(2)}+2g^{(2)}_{rr,r}+rg^{(2)}_{tt,rr}]=0
\end{array} \right.\end{equation}
The trace equation (the fourth line in the (\ref{eq2})), in
particular, provides a differential equation  with respect to the
Ricci scalar which allows to solve  the system (\ref{eq2}) at
O(2)- order\,:
\begin{equation}\label{sol}
\left\{\begin{array}{ll}
g^{(2)}_{tt}=\delta_0-\frac{\delta_1(t)e^{-r\sqrt{-\xi}}}{3\xi
r}+\frac{\delta_2(t)e^{r\sqrt{-\xi}}}{6({-\xi)}^{3/2}r}
\\\\
g^{(2)}_{rr}=\frac{\delta_1(t)[r\sqrt{-\xi}+1]e^{-r\sqrt{-\xi}}}{3\xi
r}-\frac{\delta_2(t)[\xi r+\sqrt{-\xi}]e^{r\sqrt{-\xi}}}{6\xi^2r}
\\\\
R^{(2)}=\frac{\delta_1(t)e^{-r\sqrt{-\xi}}}{r}-\frac{\delta_2(t)\sqrt{-\xi}e^{r\sqrt{-\xi}}}{2\xi
r}\end{array} \right.
\end{equation}
where $\xi=\displaystyle\frac{f_1}{6f_2}$ and $f_1$ and $f_2$ are
the expansion coefficients obtained by Taylor developing the
analytic $f(R)$ Lagrangian. Let us notice that the integration
constant $\delta_0$ is correctly dimensionless, while the two
arbitrary functions of time $\delta_1(t)$ and $\delta_2(t)$ have
respectively the dimensions of $lenght^{-1}$ and $lenght^{-2}$;
$\xi$ has the dimension $lenght^{-2}$. The functions $\delta_i(t)$
($i=1,2$)  are completely arbitrary since the differential
equation system (\ref{eq2}) contains only spatial derivatives.
Besides, the integration constant $\delta_0$ can be set to zero,
as in the theory of the potential,  since it represents an
unessential additive quantity.

With these results in mind,  the gravitational potential of a
generic analytic $f(R)$ can be obtained. In fact, the first of
(\ref{sol}) gives the second order solution in term of the metric
expansion (see the definition (\ref{definexpans})), but, as said
above, this term coincides with the gravitational potential at the
Newtonian order. In other words, we have
$g_{tt}\,=\,1+2\phi_{grav}\,=\,1+g_{tt}^{(2)}$ and then  the
gravitational potential of a fourth order gravity theory, analytic
in the Ricci scalar $R$, is
\begin{equation}\label{gravpot}
\phi_{grav}^{FOG}\,=\,\frac{K_1 e^{-r\sqrt{-\xi}}}{3\xi
r}+\frac{K_2e^{r\sqrt{-\xi}}}{6{(-\xi)}^{3/2}r}\,,
\end{equation}
with $K_1=\delta_1(t)$ and $K_2=\delta_2(t)$.

As first remark, one has to notice that the structure of the
potential, for a given $f(R)$ theory, is determined by  the
parameter $\xi$, which depends on the first and the second
derivative of the $f(R)$ function, once developed around a
particular point $R_0$.

Furthermore, one has to consider that the potential
(\ref{gravpot}) holds in the case of non-vanishing $f_2$ since we
manipulated the equations in (\ref{eq2}) dividing by such a
quantity. As matter of fact, the GR Newtonian limit cannot be
achieved directly from the solution (\ref{gravpot}) but from the
field equations (\ref{eq2}) once  the appropriate expressions in
terms of the constants $f_i$ are derived.

The solution (\ref{gravpot}) has to be  discussed in relation to
the sign of the term under the square root in  the  exponents. The
first possibility is that the sign is positive, which means that
$f_1$ and $f_2$ have opposite signature. In this case, the
solutions (\ref{sol}) and  (\ref{gravpot}) can be rewritten
introducing the  scale parameter $l=|\xi|^{-1/2}$. In particular,
considering $\delta_0\,=\,0$, the two $\delta_i(t)$ functions as
constants, $k_1\,=\,(\delta_1(t)/3)l$ and
$k_2(t)=\,(\delta_2(t)/6)l^2 $ and by introducing a  radial
coordinate $\tilde{r}$ in units of $l$, we have\,:

\begin{equation}\label{sol1}
\left\{\begin{array}{ll}
g^{(2)}_{tt}=\delta_0+\frac{\delta_1(t)l}{3}\frac{e^{-r/l}}{r/l}+\frac{\delta_2(t)l^2}{6}\frac{e^{r/l}}{r/l}=
k_1\frac{e^{-\tilde{r}}}{\tilde{r}}+k_2\frac{e^{\tilde{r}}}{\tilde{r}}
\\\\
g^{(2)}_{rr}=-\frac{\delta_1(t)l}{3}\frac{(r/l+1)e^{-r/l}}{r/l}+\frac{\delta_2(t)l^2}{6}\frac{(r/l-1)e^{r/l}}{r/l}=-
k_1\frac{(\tilde{r}+1)e^{-\tilde{r}}}{\tilde{r}}+k_2\frac{(\tilde{r}-1)e^{\tilde{r}}}{\tilde{r}}
\\\\
R^{(2)}=\frac{\delta_1(t)}{l}\frac{e^{-r/l}}{r/l}+\frac{\delta_2(t)}{2}\frac{e^{r/l}}{r/l}=\frac{3}{l^2}\biggl[k_1\frac{e^{-\tilde{r}}}
{\tilde{r}}+k_2\frac{e^{\tilde{r}}}{\tilde{r}}\biggr]\end{array}
\right.
\end{equation}
by which we can recast the gravitational potential as

\begin{equation}\label{gravpot*}
\phi_{grav}^{FOG}\,=\,\frac{k_1 e^{-\tilde{r}}}{
\tilde{r}}+\frac{k_2 e^{\tilde{r}}}{\tilde{r}}\,,
\end{equation}
which is analogous to the result  in \cite{stelle}, derived for
the theory $R+\alpha R^2+\beta R_{\mu\nu}R^{\mu\nu}$ and
coherent\symbolfootnote[5]{Let us remember that in the case of
homogeneous and isotropic spacetime, higher order curvature
invariants as $R_{\mu\nu}R^{\mu\nu}$ and
$R_{\alpha\beta\mu\nu}R^{\alpha\beta\mu\nu}$ reduce to $R^2$.}
with the results in Ref.\cite{schmidt}, obtained for higher order
Lagrangians as $f(R,\Box R)\,=\,R+\sum_{k=0}^p\,a_kR\,\Box^kR$. In
this last  case, it was demonstrated that the number of Yukawa
corrections to the gravitational potential was strictly related to
the order of the theory. However, as discussed in \cite{dirk},  it
is straightforward to show that the usual form Newton + Yukawa can
be easily achieved by Eq.(\ref{gravpot*}) through a coordinate
change.

From  (\ref{sol}) and (\ref{sol1}), one can notice that the
Newtonian limit of any analytic $f(R)$-theory is related only to
the first and second term of the Taylor expansion of the given
theory.

In other words, the gravitational potential is always
characterized by the two Yukawa corrections and only the first two
terms of the Taylor expansion of a generical $f(R)$ Lagrangian
turn out to be relevant. This is indeed a general result.

The diverging  contribution, arising from the exponential growing
mode, has to be carefully analyzed and, in particular, the
physical relevance of this term must be evaluated in relation to
the length-scale $(-\xi)^{-1/2}$. For very large $r$, (i.e.
$r>>(-\xi)^{-1/2}$),  the weak field approximation turns out to be
unphysical and the (\ref{sol}) does not hold anymore. As matter of
fact, one can obtain a modified gravitational potential which can
work as a standard Newtonian one, in the opportune limit, and
provide interesting behaviors at larger scales, even in presence
of the growing mode, once the  constants in the (\ref{gravpot})
have been opportunely adjusted. Such a potential, once the growing
exponential term is settled to zero, reproduce the Yukawa-like
gravitational potential, phenomenologically introduced by Sanders
\cite{sanders} to explain the flat rotation curves of spiral
galaxies without dark matter.

Besides, Yukawa-like corrections to the gravitational potential
have been suggested in several approaches. For example, an
interesting proposal is a model describing the gravitational
interaction between dark matter and baryons. This points out that
 the interaction suppressed on small subgalactic scales can be
described by means of a Yukawa contribution added to the standard
Newtonian potential. Such a behavior is effectively suggested by
the  observations of the inner rotation curves of  low-mass
galaxies and provides a natural scenario in which to interpret the
cuspy profile of dark matter halos observed in N-body simulations
\cite{piazza}.

It is important to stress that the result we have obtained here is
coherent with other  calculations. In fact, since the Taylor
expansion of an exponential potential is a power law series, it is
not surprising to obtain a power law correction to the Newtonian
potential \cite{noi-mnras} when a less rigorous approach is
considered in order to calculate the weak field limit of a generic
$f(R)$-theory. In particular, perturbative calculations will
provide effective potentials which can be recovered by means of an
appropriate approximation from the general case  (\ref{gravpot*}).

Let us now consider now the opposite case in which the sign of
$\xi$ is negative and, as a consequence, the two Yukawa
corrections in (\ref{sol1}) are complex numbers.

Since the form of $g_{tt}$, the  gravitational potential
(\ref{gravpot*}) turns out to be\,:
\begin{equation}\label{gravpottrig}
\phi_{grav}^{FOG}\,=\,\frac{k_1 e^{-\imath\tilde{r}}}{
\tilde{r}}+\frac{k_2 e^{\imath\tilde{r}}}{\tilde{r}}\,,
\end{equation}
which can be recast as
\begin{equation}\label{gravpottrig1}
\phi_{grav}^{FOG}\,=\,\frac{1}{\tilde{r}}\left[(k_1+k_2)\cos{{\tilde{r}}}+i(k_2-k_1)\sin{\tilde{r}}\right]\,.
\end{equation}
Such a gravitational potential, which could be discarded as a
non-physically relevant, has the  property to satisfy the
Helmholtz equation, $\nabla^2\phi+k^2\phi=4\pi G\rho$, where
$\phi$ is the gravitational potential and $\rho$ is a real
function acting both as matter and the antimatter density. As
discussed in \cite{bartlett}, $Re\,[\phi_{grav}^{FOG}]$ can be
addressed as a classically modified Newtonian potential corrected
by  a Yukawa factor while $Im\,[\phi_{grav}^{FOG}]$ could have
significant implications for quantum mechanics. In particular,
this term can provide an astrophysical, and in our case even
theoretically well founded, origin for the puzzling decay
$K_L\rightarrow \pi^+\pi^-$ whose phase is related to an imaginary
potential in the kaon mass matrix. Of course, these
considerations,  at this level, are only speculative, nevertheless
it could be worth  taking them into account for further
investigations.

Let us now consider the  system (\ref{sys1}) up to the third order
contributions. The first important issue is that, at this order,
one has to consider even the off-diagonal equation

\begin{equation}\label{off-d}
f_1g^{(2)}_{rr,t}+2f_2rR^{(2)}_{,tr}=0\,,
\end{equation}
which relate the time derivative of the Ricci scalar to the time
derivative of $g^{(2)}_{rr}$. From this relation, it is possible
to draw a  relevant consideration. One can deduce that, if the
Ricci scalar depends on time so it is for the metric components
and even the gravitational potential turns out to be influenced.
This result agrees with the analysis provided in \cite{artspher}
where a complete description of the weak field limit of fourth
order gravity has been provided in term of the dynamical evolution
of the Ricci scalar. In that paper, it was demonstrated that if
one supposes a time independent Ricci scalar, static spherically
symmetric solutions are allowed. Eq.(\ref{off-d}) confirms this
result and provides the formal theoretical explanation of such a
behavior. In particular, together with the (\ref{sol1}), it
suggests that if one considers the problem at a lower level of
approximation (i.e. the second order) the background spacetime
metric can have static solutions according to the Birkhoff
theorem; this is no more verified when the problem is faced with
approximations of higher order. In other words, the debated issue
to prove the validity of the Birkhoff theorem in the  higher order
theories of gravity, finds here its physical answer. In
\cite{artspher} and here, the validity of this theorem is
demonstrated for $f(R)$ theories only when the Ricci scalar is
time independent or, in addition, when the Newtonian limit
solutions are investigated up to the second order of approximation
in term of a $v/c$ expansion of the metric coefficients.
Therefore, the Birkhoff theorem does not represent a general
feature in the case of fourth order gravity but, on the other
hand, in the limit of small velocities and weak fields (which is
enough to deal with the Solar System gravitational experiments),
one can assume that the gravitational potential is effectively
time independent according to (\ref{sol}) and (\ref{gravpot}).

The above results fix a fundamental difference between GR and
fourth order gravity theories. While in GR a spherically symmetric
solution represents a stationary and static configuration
difficult to be related to a cosmological background evolution,
this is no more true in the case of higher order gravity. In the
latter case, a spherically symmetric background can have
time-dependent evolution together with the radial dependence. In
this sense, a relation between a spherical solution and the
cosmological Hubble flow can be easily achieved.

The subsequent step concerns the analysis of the system
(\ref{sys1}) up to the $O(4)$ order. Such an analysis provides the
solutions, in term of $g_{tt}^{(4)}$,  the right order for the
post-Newtonian parameters. Unfortunately, at this order of
approximation, the system turns out to be too much involuted and a
general solution is not possible.

From Eqs. (\ref{sys1}), one can notice that the general solution
is characterized only by the first three orders of the $f(R)$
expansion. Such a result is in agreement with the $f(R)$
reconstruction which can be induced by the post-Newtonian
parameters adopting a scalar-tensor analogy (for details see
\cite{ppn-noi,ppn-noi-bis}).

However, although we cannot achieve a complete description,  an
approximate estimation  of the post-Newtonian parameter $\gamma$
can be obtained recurring to the $O(2)$ evaluation of the metric
coefficients in the vacuum case.

It is important to notice that, since (\ref{sol}) suggests a
modified gravitational potential (with respect to the standard
Newtonian one) as a general solution of  analytic $f(R)$ gravity
models, there is no reason to ask for a post-Newtonian description
for these theories. In fact, as previously said, the
post-Newtonian analysis presupposes to evaluate deviations from
the Newtonian potential at a higher than second order
approximation in term of the quantity $v/c$. Thus, if the
gravitational potential deduced from a given $f(R)$ theory of
gravity is a  general function of the radial coordinate,
displaying a Newtonian behavior only in a certain regime (or in a
given range of the radial coordinate), it could be meaningless to
develop a general post-Newtonian formalism as in GR
\cite{will,nordvedt}. Of course, by a proper expansion of the
gravitational potential for small values of the radial coordinate,
and only in this limit, one can develop an analogous of the
post-Newtonian limit for these theories with respect to the
Newtonian behavior and estimate the deviations from it.

In order to have an effective estimation  of the post-Newtonian
parameter $\gamma$, we can proceed in the following way. Expanding
$g_{tt}$ and $g_{rr}$, obtained at the second order in
(\ref{sol1}) with respect to the dimensionless coordinate
$\tilde{r}$, one has\symbolfootnote[6]{In this case the symbol
$O[2]$ is referred to higher than first order contributions the
dimensionless coordinate $\tilde{r}.$}

\begin{eqnarray}\nonumber
g^{(2)}_{tt}=(k_2-k_1)+\frac{k_1+k_2}{\tilde{r}}+\frac{k_1+k_2}{2}\tilde{r}+O[2]\,,
\\\label{taylorppn}\\
g^{(2)}_{rr}=-\frac{k_1+k_2}{\tilde{r}}+\frac{k_1+k_2}{2}\tilde{r}+O[2]\,,
\nonumber
\end{eqnarray}
where, clearly, $k_1+k_2=GM$ and $k_1=k_2$ in the standard case.
When $\tilde{r}\rightarrow 0$ (i.e. when the coordinate
$r<<\sqrt{-\xi}$) the linear and the higher than first order terms
are vanishingly small and  only the first Newtonian term survives.
Since the post-Newtonian parameter $\gamma$ is strictly related to
the coefficients of the $1/r$ term into the expressions of
$g_{tt}$ and $g_{rr}$, actually one can obtain an effective
estimation  of this quantity confronting the coefficients of the
Newtonian terms relative to both the expressions in
(\ref{taylorppn}). Being $\gamma\,=\,1$ in GR, the difference
between these two coefficients  gives the effective deviation from
the GR expectation value.

It is easy to derive that a generic fourth order gravity theory
provides a post-Newtonian parameter $\gamma$ which is consistent
with the GR prescription $(\gamma\,=\,1)$ if $k_1\,=\,k_2$.
Conversely, deviations from such a behavior can be accommodated by
tuning the relation between the two integration constants $k_1$
and $k_2$. This is equivalent to adjust the form of the $f(R)$
theory in such a way to obtain the right GR limit, and then the
Newtonian potential. This result agrees with the viewpoint that
asks for the recovering of GR behavior from generic $f(R)$
theories in the post-Newtonian limit \cite{zhang,sotiriou}. This
is particularly true when the $f(R)$ Lagrangian behaves, in the
weak field and small velocities regime, as the Hilbert-Einstein
Lagrangian.

On the other side, if deviations from these regime are observed, a
$f(R)$ Lagrangian, built up with a third order polynomial in the
Ricci scalar, can be suitable  to interpret such a behavior (see
\cite{ppn-noi-bis}).

Actually, the degeneracy regarding the integration constants can
be partially broken once a complete post-Newtonian
parameterization is developed in presence of matter. In such a
case,  the integration constants  remain constrained by the
Boltzmann-Vlasov equation which describes the conservation of
matter at these scales \cite{binney}.

Up to now, the  discussion has been developed without  any gauge
choice. In order to overcome the difficulties related to the
nonlinearities of calculations, we can work considering some gauge
choice  obtaining  less general solutions for the metric entries.
A natural choice  is represented by the conditions (\ref{PPNobj})
which  coincide with  the standard post-Newtonian gauge
\begin{eqnarray}
 h_{jk,}{}^{k} - \frac{1}{2} h_{,j} &=& {O}(4),
\nonumber \\
h_{0k,}{}^{k} - \frac{1}{2} h^{k}{}_{k,0} &=& {O}(5)\,,
\end{eqnarray}
where $h_{\mu\nu}$ accounts for deviations from the Minkowski
metric ($g_{\mu\nu}\,=\,\eta_{\mu\nu}+h_{\mu\nu}$). In this case
the Ricci curvature tensor becomes\footnote{We have indicated with
the subscript $_{hg}$ the harmonic gauge variables.}

\begin{equation}
\left\{\begin{array}{ll} R_{tt|_{hg}}&\simeq
R^{(2)}_{tt|_{hg}}+R^{(4)}_{tt|_{hg}}
\\\\
R_{rr|_{hg}}&\simeq R^{(2)}_{rr|_{hg}}\\\\
\end{array} \right.
\end{equation}
where

\begin{equation}
\left\{\begin{array}{ll}
R^{(2)}_{tt|_{hg}}=\frac{rg^{(2)}_{tt,rr}+2g^{(2)}_{tt,r}}{2r}
\\\\R^{(4)}_{tt|_{hg}}=\frac{rg^{(4)}_{tt,rr}+2g^{(4)}_{tt,r}+r[g^{(2)}_{rr}g^{(2)}_{tt,rr}-g^{(2)}_{tt,tt}-{g^{(2)}
_{tt,rr}}^2]}{2r}\\\\
R^{(2)}_{rr|_{hg}}=\frac{rg^{(2)}_{rr,rr}+2g^{(2)}_{rr,r}}{2r}\\\\R^{(2)}_{\theta\theta|_{hg}}=R^{(2)}_{\phi\phi|_{hg}}=0
\end{array} \right.
\end{equation}
while the Ricci scalar expressions at the $O(2)$ and $O(4)$ orders
read

\begin{equation}
\left\{\begin{array}{ll}
R^{(2)}_{|hg}=\frac{rg^{(2)}_{tt,rr}+2g^{(2)}_{tt,r}-rg^{(2)}_{rr,rr}-2g^{(2)}_{rr,r}}{2r}\\\\
R^{(4)}_{|hg}=\frac{rg^{(4)}_{tt,rr}+2g^{(4)}_{tt,r}+r[g^{(2)}_{rr}g^{(2)}_{tt,rr}-g^{(2)}_{tt,tt}-{g^{(2)}
_{tt,rr}}^2]-g^{(2)}_{tt}[rg^{(2)}_{tt,rr}+2g^{(2)}_{tt,r}]-g^{(2)}_{rr}[rg^{(2)}_{rr,rr}+2g^{(2)}_{rr,r}]}{2r}
\end{array} \right..
\end{equation}
The gauge choice does not affect the Christoffel. Thus, by solving
the  system (\ref{sys1}), with the simplification induced by the
gauge, one obtains

\begin{equation}\label{armonic}
\left\{\begin{array}{ll}g_{tt_{|_{hg}}}(t,r)=1+\frac{k_1}{r}+\frac{k_2}{r^2}+\frac{k_3 \log r}{r}\\\\
g_{rr_{|_{hg}}}(t,r)=1+\frac{k_4}{r}\end{array} \right.
\end{equation}
where the constants $k_1$, $k_4$ are relative to the $O(2)$ order
of approximation, while $k_2$ and $k_3$ are related to the $O(4)$
order. The Ricci scalar is zero both at  $O(2)$ and at $O(4)$
approximation orders.

Eqs.(\ref{armonic}) suggest some interesting remarks. It is easy
to check that the GR prescriptions are immediately recovered for
$k_1\,=\,k_4$ and $k_2\,=\,k_3\,=0$.  The $g_{rr}$ component
displays only the second order term, as required by a GR-like
behavior, while the $g_{tt}$ component shows also the fourth order
corrections which determine the second post-Newtonian parameter
$\beta$ \cite{will}. It has to be stressed here that a full
post-Newtonian formalism requires to take into account  matter in
the system ({\ref{sys1}}): the presence of matter links the second
and fourth order contributions in the metric coefficients
\cite{will}.

\section{The Post-minkowskian approximation}

In the previous section we have developed a general analytic
procedure to deduce the Newtonian and the post-Newtonian limit of
$f(R)$-gravity in absence of matter or far from  matter sources.
Here we want to discuss a different limit of these theories,
pursued when the small velocity assumption is relaxed and only the
weak field approximation is retained. This situation is  related
to the Minkowski limit of the underlying gravity theory as well as
the discussion of the previous section was related to the
Newtonian one. In order to develop such an analysis, we can
reasonably resort to the metric (\ref{me}), considering the
gravitational potentials $A(t,r)$ e $B(t,r)$ in the suitable form

\begin{equation}
\left\{\begin{array}{ll}A(t,r)=1+a(t,r)\\\\
B(t,r)=1+b(t,r)\end{array} \right.
\end{equation}
with $a(t,r), b(t,r)\ll 1$. Let us now perturb the field equations
(\ref{fe1}) considering, again,  the Taylor expansion
(\ref{sertay}) for a generic $f(R)$ theory. For the vacuum case
($T_{\mu\nu}=0$), at the first order with respect to $a$ e $b$, it
is
\begin{equation}\label{eq4}
\left\{\begin{array}{ll}f_0=0\\\\
f_1\biggl\{R^{(1)}_{\mu\nu}-\frac{1}{2}g^{(0)}_{\mu\nu}R^{(1)}\biggr\}+\mathcal{H}^{(1)}_{\mu\nu}=0\end{array}
\right.
\end{equation}
where

\begin{eqnarray}\mathcal{H}^{(1)}_{\mu\nu}=-f_2\biggl\{R^{(1)}_{,\mu\nu}-{\Gamma^{(0)}}^{\rho}_{\mu\nu}R^{(1)}_{,\rho}-g^{(0)}_{\mu\nu}\biggl[
{g^{(0)\rho\sigma}}_{,\rho}R^{(1)}_{,\sigma}+g^{(0)\rho\sigma}R^{(1)}_{,\rho\sigma}+
g^{(0)\rho\sigma}\ln\sqrt{-g}^{(0)}_{,\rho}R^{(1)}_{,\sigma}\biggr]\,.
\biggr\}
\end{eqnarray}
In this approximation, the Ricci scalar turns out to be  zero
while  the derivatives, in the previous relations, are calculated
at $R=0$.

Let us now consider the limit for large $r$, i.e. we study the
problem far from the source of the gravitational field. In such a
case Eqs. (\ref{eq4}) become
\begin{equation}\label{eq5}
\left\{\begin{array}{ll}\frac{\partial^2a(t,r)}{\partial r^2}-\frac{\partial^2b(t,r)}{\partial t^2}=0\\\\
f_1\biggl[a(t,r)-b(t,r)\biggr]-8f_2\biggl[\frac{\partial^2b(t,r)}{\partial
r^2}+\frac{\partial^2a(t,r)}{\partial
t^2}-2\frac{\partial^2b(t,r)}{\partial
t^2}\biggr]=\Psi(t)\end{array} \right.
\end{equation}
where $\Psi(t)$ is a generic time-dependent function.
Eqs.(\ref{eq5}) are two coupled wave equations in term of the two
functions $a(t,r)$ and $b(t,r)$. Therefore, we can ask for a
wave-like solutions for the gravitational potentials $a(t,r)$ and
$b(t,r)$
\begin{equation}
\left\{\begin{array}{ll}a(t,r)=\int\frac{d\omega
dk}{2\pi}\tilde{a}(\omega,k)e^{i(\omega
t-kr)}\\\\
b(t,r)=\int\frac{d\omega dk}{2\pi}\tilde{b}(\omega,k)e^{i(\omega
t-kr)}\end{array} \right.
\end{equation}
and substituting these into the (\ref{eq5}). In order to simplify
the calculations, we can set $\Psi(t)=0$ since, as said, this is
an arbitrary time function.  Eqs.(\ref{eq5}) are satisfied if

\begin{equation}
\left\{\begin{array}{ll} \begin{array}{ccc}
  \tilde{a}(\omega,k)=\tilde{b}(\omega,k)\,, & \,\,\, & \omega=\pm k \\
  & & \\
  \tilde{a}(\omega,k)=\biggl[1-\frac{3\xi}{4k^2}\biggr]\tilde{b}(\omega,k)\,, & \,\,\, & \omega=\pm\sqrt{k^2-\frac{3\xi}{4}} \\
  \end{array}
\end{array} \right.
\end{equation}
where, as before, ${\displaystyle \xi=\frac{f_1}{6f_2}}$.  In
particular, for $f_1=0$ or $f_2=0$ one  obtains solutions with a
dispersion relation $\omega=\pm k$. In other words, for $f_i \neq
0$ ($i\,=\,1,2$), that is in the case of non-linear $f(R)$, the
above dispersion relation  suggests that massive modes are in
order. In particular, for $\xi<0$,  the mass of the graviton is
${\displaystyle m_{grav}\,=\,-\frac{3\xi}{4}}$ and, coherently, it
is obtained for a modified real gravitational potential. As matter
of fact, a gravitational potential deviating from the Newtonian
regime induces a massive degree of freedom into the particle
spectrum of the gravity sector with interesting perspective for
the detection and the production of gravitational waves
\cite{CCD}. It has to be remarked that the presence of massive
gravitons in the wave spectrum of higher order gravity is a well
known result since the paper of \cite{stelle}. Nevertheless it is
our opinion that this issue has been always considered under a
negative perspective and has been not sufficiently investigated.
Furthermore, if $\xi>0$, even the solution

\begin{equation}
\left\{\begin{array}{ll}a(\tilde{t},\tilde{r})=(a_0+a_1\tilde{r})e^{\pm\frac{\sqrt{3}}{2}\tilde{t}}\\\\
b(\tilde{t},\tilde{r})=(b_0+b_1\tilde{t})\cos\biggl[\frac{\sqrt{3}}{2}\tilde{r}\biggr]+(b'_0+b'_1\tilde{t})\sin\biggl[\frac{\sqrt{3}}{2}\tilde{r}\biggr]+b''_0+
b''_1\tilde{t}\end{array} \right.
\end{equation}
with $a_0$, $a_1$, $b_0$, $b_1$, $b'_0$, $b'_1$, $b''_0$, $b''_1$
constants is admitted. The variables $\tilde{r}$ and $\tilde{t}$
are expressed in units of $\xi^{-1/2}$. In the post-Minkowskian
approximation, as expected, the gravitational field propagates by
means of wave-like solutions. This result suggests that
investigating the gravitational waves behavior of fourth order
gravity can represent an interesting issue where a new
phenomenology (massive gravitons) has to be seriously taken  into
account. Besides, such massive degrees of freedom could be a
realistic and testable candidate for cold dark matter, as
discussed in \cite{massivegrav}.

\section{Conclusions}

In this paper, we have developed a general analytic approach to
deal with the weak field and small velocity limit (Newtonian
limit) of a generic $f(R)$ gravity theory. The scheme  can be
adopted also to correctly calculate the post-Newtonian parameters
of such theories without any redefinition of the degrees of
freedom  by some scalar field leading to the so called O'Hanlon
Lagrangian \cite{ohanlon}. In fact, considering this latter
approach, we get a Brans-Dicke like theory with vanishing kinetic
term and then the post-Newtonian parameter $\gamma$ results
$\gamma=1/2$ and not $\gamma\sim 1$ as observed. This result is
misleading in the weak field limit. In the approach presented
here, we do not need any change from the Jordan to the Einstein
frame \cite{ppn-noi-bis,sotiriou2}. Apart the possible
shortcomings related to  non-correct changes of variables, any
$f(R)$ theory can be rewritten  as a scalar-tensor one or  an
ideal fluid, as shown in \cite{nodi,nodicap1,nodicap2}. In those
papers, it has been demonstrated that such different
representations give rise to physically non-equivalent theories
and then also the Newtonian and post-Newtonian approximations have
to be handled very carefully because the results could not be
equivalent. In fact, the further geometric degrees of freedom of
$f(R)$ gravity (with respect to GR), the scalar field and the
ideal fluid have weak field behaviors strictly depending on the
adopted gauge which could not be equivalent or difficult to
compare. In order to circumvent these possible sources of
shortcomings, one should states the frame (Jordan or Einstein) at
the very beginning and then remain in such a frame along all the
calculations up to the final results. Adopting this procedure,
arbitrary limits and non-compatible results should be avoided.

In this paper, we have considered the Taylor expansion of a
generic $f(R)$ theory,  obtaining general solutions in term of the
metric coefficients up to the third order of approximation when
matter is neglected. In particular, the solution relative to the
$g_{tt}$ metric component gives the gravitational potential which
is corrected with respect to the Newtonian one of $f(R)=R$. The
general gravitational potential is given by a couple of
Yukawa-like terms, combined with the Newtonian potential, which is
effectively achieved at small distances. In relation to the sign
of the characteristic coefficients entering the $g_{tt}$
component, one can obtain real or complex solutions. In both
cases, the resulting gravitational potential   has physical
meanings. This degeneracy could be removed once standard matter is
introduced into dynamics.

The complete analysis allows to obtain direct information on the
post-Newtonian formalism: the post-Newtonian parameters can be
fully characterized considering the integration constants in the
gravitational potential. Nevertheless this study is beyond the aim
of this paper and will be developed in a forthcoming research
project.

Furthermore, it has been shown that the Birkhoff theorem is not a
general result for $f(R)$-gravity. This is a fundamental
difference between GR and fourth order gravity. While in GR a
spherically symmetric solution is, in any case,  stationary and
static, here time-dependent evolution can be achieved depending on
the order of perturbations.

Finally, we have discussed the differences between the
post-Newtonian and the post-Minkoskian limit in $f(R)$ gravity.
The main result of such an investigation is the presence of
massive degrees of freedom in the  spectrum of gravitational waves
which are strictly related to the modifications occurring into the
gravitational potential. This occurrence could constitute an
interesting opportunity for the detection and investigation of
gravitational waves.

\section{Appendix}

In this appendix, we show that the  harmonic gauge  can be
suitably reduced to the form (\ref{PPNobj}). Such a gauge is
usually characterized by  the condition
$g^{\sigma\tau}\Gamma^{\mu}_{\sigma\tau}=0$. For $\mu=0$ one has

\begin{equation}\label{gau1}
2g^{\sigma\tau}\Gamma^{0}_{\sigma\tau}\approx {g^{(2)}}^{0,0}_{0}
-2{g^{(3)}}^{0,m}_{m}+{g^{(2)}}^{m,0}_{m}=0\,,
\end{equation} and for
$\mu=i$

\begin{equation}\label{gau2}
2g^{\sigma\tau}\Gamma^{i}_{\sigma\tau}\approx{g^{(2)}}^{0,i}_{0}
+2{g^{(2)}}^{mi}_{,m}-{g^{(2)}}^{m,i}_{m}=0\,.
\end{equation}
Differentiating Eq.(\ref{gau1}) with respect to $x^0$, $x^j$ and
(\ref{gau2}) and with respect to $x^0$, one obtains

\begin{equation}\label{gau3}
{g^{(2)}}^{0}_{0,00}-2{g^{(3)}}^{m}_{0,0m}+{g^{(2)}}^{m}_{m,00}=0\,,
\end{equation}

\begin{equation}\label{gau4}
{g^{(2)}}^{0}_{0,0j}-2{g^{(3)}}^{m}_{0,jm}+{g^{(2)}}^{m}_{m,0j}=0\,,
\end{equation}

\begin{equation}\label{gau5}
{g^{(2)}}^{0}_{0,0i}+2{g^{(2)}}^{m}_{i,0m}-{g^{(2)}}^{m}_
{m,0i}=0\,.
\end{equation}
On the other side, combining Eq.(\ref{gau4}) and Eq.(\ref{gau5}),
we get

\begin{equation}\label{gau6}
{g^{(2)}}^{m}_{m,0i}-{g^{(2)}}^{m}_{i,0m}-{g^{(3)}}^{m}_
{0,mi}=0\,.
\end{equation}

Finally, differentiating Eq.(\ref{gau2}) with respect to $x^j$,
one has\,:

\begin{equation}\label{gau7}
{g^{(2)}}^{0}_{0,ij}+2{g^{(2)}}^{m}_{i,jm}-{g^{(2)}}^{m}_ {m,ij}=0
\end{equation}
and redefining indexes as $j\rightarrow i$, $i\rightarrow j$ since
these are mute indexes, we get

\begin{equation}\label{gau8}
{g^{(2)}}^{0}_{0,ij}+2{g^{(2)}}^{m}_{j,im}-{g^{(2)}}^{m}_
{m,ij}=0\,.
\end{equation}
Combining  Eq.(\ref{gau7}) and  Eq.(\ref{gau8}), we obtain

\begin{equation}\label{gau9}
{g^{(2)}}^{0}_{0,ij}+{g^{(2)}}^{m}_{i,jm}+{g^{(2)}}^{m}_{j,im}-{g^{(2)}}^{m}_
{m,ij}=0\,.
\end{equation}
The relations (\ref{gau3}), (\ref{gau6}), (\ref{gau9})  guarantee
the viability of (\ref{PPNobj}).


\begin{thebibliography}{99}

\bibitem{sneIa} A.G. Riess {\it et al.} Astron. J. {\bf 116}, 1009 (1998); A.G. Riess {\it et al.} Astroph. Journ. {\bf 607}, 665
(2004); S. Perlmutter {\it et al.} Astrophys. J {\bf 517}, 565
(1999); Astron. Astrophys. {\bf 447}, 31 (2006).

\bibitem{lss}
S. Cole {\it et al.}, Mon. Not. Roy. Astron. Soc. {\bf 362}, 505
(2005).

\bibitem{cmbr}
D.N. Spergel {\it et al.} Astrophys. J. Suppl. {\bf 148}, 175
(2003); D.N. Spergel {\it et al.} arXiv: astro-ph/0603449.

\bibitem{f(R)-noi}
S. Capozziello, Int. J. Mod. Phys. {\bf D 11}, 483, (2002); S.
Capozziello, S. Carloni, A. Troisi, Rec. Res. Develop. Astron.
Astrophys. {\bf 1}, 625 (2003), arXiv:astro\,-\,ph/0303041; S.
Capozziello, V.F. Cardone, S. Carloni, A. Troisi, Int. J. Mod.
Phys. {\bf D}, 12, 1969 (2003).

\bibitem{f(R)-cosmo}
S.~M.~Carroll, V.~Duvvuri, M.~Trodden and M.~S.~Turner, Phys.\
Rev.\ D {\bf 70}, 043528 (2004); S. Nojiri, S.D. Odintsov, Phys.
Lett. {\bf B 576}, 5, (2003); S. Nojiri, S.D. Odintsov, Phys. Rev.
{\ bf D 68}, 12352, (2003); S.~Capozziello, V.~F.~Cardone and
A.~Troisi, Phys.\ Rev.\ {\bf D 71}, 043503 (2005); S. Carloni,
P.K.S. Dunsby, S. Capozziello, A. Troisi, Class. Quant. Grav. {\bf
22}, 4839 (2005).

\bibitem{palatini}
D.N. Vollick, Phys. Rev. {\bf D 68}, 063510, (2003); X.H. Meng, P.
Wang, Class. Quant. Grav., {\bf 20}, 4949, (2003); E.E. Flanagan,
Phys. Rev. Lett. {\bf 92}, 071101, (2004); E.E. Flanagan, Class.
Quant. Grav., {\bf 21}, 417, (2004); X.H. Meng, P. Wang, Class.
Quant. Grav., {\bf 21}, 951, (2004); G.M. Kremer, D.S.M. Alves,
gr\,-\,qc/0404082, 2004; G. Allemandi, A. Borowiec, M.
Francaviglia, Phys. Rev. {\bf D 70}, 103503 (2004); S.
Capozziello, V.F. Cardone, M. Francaviglia, {\it Gen. Rel. Grav.}
{\bf 38}, 711 (2006).


\bibitem{GRGreview}
S. Capozziello and M. Francaviglia, 0706.1146 [astro\,-\,ph]
(2007).

\bibitem{OdintsovLadek}
S.Nojiri and S.D. Odintsov, {\it Int. J. Meth. Mod. Phys.} {\bf
4}, 115 (2007).

\bibitem {farhoudi}M. Farhoudi, \textsl{Gen. Relativ. Grav.}\ \textbf{38},
1261 (2006).

\bibitem{starobinsky}
A. A. Starobinsky, Phys. Lett. {\bf B} 91, 99 (1980).

\bibitem{stelle}
K.S. Stelle, Gen. Rev. Grav. {\bf 9}, 343, (1978).

\bibitem{noi-mnras} S. Capozziello, V.F. Cardone, A. Troisi, JCAP {\bf 0608}, 001
(2006);\\
S. Capozziello, V.F. Cardone, A. Troisi, Mon.\ Not.\ Roy.\
Astron.\ Soc.\  {\bf 375}, 1423 (2007).

\bibitem{salucci}
C. Frgerio Martins and P. Salucci, to appear in MNRAS, arXiv:
astro\,-\,ph/0703243 (2007).

\bibitem{anderson}
J.D. Anderson et al. {\it Phys. Rev. Lett.} {\bf 81}, 2858
(1998);\\
J.D. Anderson et al. {\it Phys. Rev.} {\bf D 65}, 082004 (2002).

\bibitem{bertolami}
O. Bertolami et al. arXiv: 0704.1733 [gr\,-\,qc] (2007).


\bibitem{olmo}
G.J. Olmo, Phys. Rev. Lett. {\bf 95}, 261102 (2005)

\bibitem{ppn-noi}
S. Capozziello, A. Troisi, Phys. Rev. {\bf D 72}, 044022, (2005);


\bibitem{ppn-ok}
T. P. Sotiriou,  
  Gen.\ Rel.\ Grav.\  {\bf 38}, 1407 (2006); G.~Allemandi, M.~Francaviglia, M.~L.~Ruggiero and
A.~Tartaglia, 
  Gen.\ Rel.\ Grav.\  {\bf 37}, 1891 (2005); S. Nojiri and S.D.
  Odintsov, ArXiv: 0707.1941, [hep\,-\,th] (2007).


\bibitem{ppn-no}
T. Chiba, Phys. Lett. {\bf B 575}, 1 (2005); T. Clifton and J.D.
Barrow, {\it Phys. Rev.} {\bf D 72}, 103005 (2005); V. Faraoni,
arXiv: gr-qc:0607016; A.L. Erickcek, T.L. Smith, M. Kamionkownski,
arXiv: astro-ph/0610483; X.~H.~Jin, D.~J.~Liu and X.~Z.~Li,
arXiv:astro-ph/0610854.  
T. Chiba, A.L. Erickcek, T.L. Smith,  arXiv: astro-ph/0611867
(2006).

\bibitem{faraonithomas}
T. P. Sotiriou, V. Faraoni, S. Liberati, arXiv: 0707.2748
[gr\,-qc] (2007).

\bibitem{dirk}
A. Stabile, D. Puetzfeld, S. Capozziello, in preparation(2007).

\bibitem{artspher}
S. Capozziello, A. Stabile, A. Troisi, in preparation (2007).

\bibitem{noether}
 S. Capozziello, A. Stabile, A. Troisi {\it Class. Quant. Grav.}
{\bf 24},  2153  (2007).

\bibitem{multamaki}
T. Multam\"aki, I. Vilja,  {\it Phys. Rev.} {\bf D 74}, 064022 (2006);\\
T. Multam\"aki, I. Vilja,  arXiv:astro-ph/0612775 (2006).


\bibitem{will}
C.M  Will, {\it Theory and experiment in gravitational physics},
(Cambridge University Press, Cambridge, U.K.; New York, U.S.A.,
1993), 2nd edition; C.~M.~Will, Living Rev.\ Rel.\  {\bf 4}, 4
(2001).

\bibitem{schmidt}
I.~Quandt and H.~J.~Schmidt,  Astron.\ Nachr.\  {\bf 312}, 97
(1991);\\ H. J. Schmidt, gr-qc/0602017 (2006).


\bibitem{sanders}
 R.H. Sanders, {\it Ann. Rev. Astron. Astroph.} {\bf 2}, 1 (1990).

\bibitem{piazza}
F.~Piazza and C.~Marinoni,  Phys.\ Rev.\ Lett.\  {\bf 91}, 141301
(2003).

\bibitem{bartlett}
D. F. Bartlett, Am. J. Phys. {\bf 62}, 8, (1994);\\
D.~F.~Bartlett,Int.\ J.\ Mod.\ Phys.\  A {\bf 16S1B}, 680 (2001).

\bibitem{ppn-noi-bis}
S. Capozziello, A. Stabile, A. Troisi, Mod.\ Phys.\ Lett.\ {\bf A
21}, 2291 (2006).

\bibitem{nordvedt}
K.~Nordtvedt, Phys.\ Rev.\  {\bf 169}, 1017 (1968);

\bibitem{zhang}
P. Zhang, arXiv:astro-ph/0701662 (2007).

\bibitem{sotiriou}
T.~P.~Sotiriou and E.~Barausse,
  Phys.\ Rev.\  D {\bf 75}, 084007 (2007)



\bibitem{multamaki06}
T. Multam$\ddot{a}$ki, I. Vilja, arXiv: astro-ph/0612775.


\bibitem{binney}
Binney, J., Tremaine, S. 1987, {\it Galactic dynamics}, Princeton
University Books, Princeton (USA).

\bibitem{CCD}
S. Capozziello, Ch. Corda. M. De Laurentis, {\it Mod. Phys. Lett.}
{\bf. A 22}, 1097 (2007).

\bibitem{massivegrav}
S.L. Dubovsky, P.G. Tiyakov, and I.I. Tkachev, {\it Phys. Rev.
Lett.} {\bf 94}, 181102 (2005).

\bibitem{ohanlon}
J. O'Hanlon, {\it Phys. Rev.Lett.} {\bf 29}, 137 (1972).

\bibitem{sotiriou2}
T. P. Sotiriou, {\it Class. Quant. Grav.} {\bf 23}, 5117 (2006).

\bibitem{nodi}
S. Nojiri and S.D. Odintsov, {\it  Phys. Rev.} {\bf D 74}, 086005
(2006).

\bibitem{nodicap1}
S. Capozziello, S. Nojiri and S.D. Odintsov, {\it  Phys. Lett.}
{\bf B 634}, 93 (2006).

\bibitem{nodicap2}
S. Capozziello, S. Nojiri, S.D. Odintsov, and A. Troisi {\it Phys.
Lett.} {\bf B 639}, 135 (2006).












\end{thebibliography}
\end{document}